\documentclass[letter,12pt]{article}

\usepackage[T1]{fontenc}
\usepackage[utf8]{inputenc}
\usepackage{lmodern}

\usepackage[english]{babel}
\usepackage{csquotes}

\usepackage{authblk}

\usepackage{hyperref}
\hypersetup{
	colorlinks,
	linkcolor={black},
	citecolor={black},
	urlcolor={blue}
}
\usepackage{float}
\usepackage{acronym}
\usepackage{setspace}
\usepackage{etoolbox}

\usepackage[style=apa,natbib,sortcites,backend=biber,date=year]{biblatex}
\AtEveryBibitem{\clearfield{month}}
\AtEveryBibitem{\clearfield{day}}
\DeclareLanguageMapping{british}{british-apa}

\newtoggle{final}
\newtoggle{blind}

\togglefalse{final}
\togglefalse{blind}

\iftoggle{blind}{
\usepackage{draftwatermark}
\SetWatermarkText{blinded \today}
\SetWatermarkLightness{0.9}
\SetWatermarkScale{0.3}
}{}

\newcommand{\blindurl}[1]{\iftoggle{blind}{[URL blinded]}{\url{#1}}}

\iftoggle{final}{%
    \doublespacing
}{
	\usepackage{draftwatermark}
	\SetWatermarkText{DRAFT}
	\SetWatermarkLightness{0.9}
	\SetWatermarkScale{0.3}
}

\acrodef{AEA}{American Economic Association}
\acrodef{LDI}{Labor Dynamics Institute}
\acrodef{DCAP}{Data and Code Availability Policy}

\title{Teaching for large-scale Reproducibility Verification}

\iftoggle{blind}{
    \newcommand{\university}{{\it ABC University}}
	\author[1]{AUTHORS 1, 2, 3, 4, 5}
	\affil[1]{\university{}}
}%
{	
\newcommand{\university}{Cornell University}
	
\author[1,2]{Lars Vilhuber}
\author[1]{Hyuk Harry Son}
\author[1]{Meredith Welch}
\author[1]{David N. Wasser}
\author[1]{Michael Darisse}
\affil[1]{\university{}}
\affil[2]{\it \footnotesize As per Editor's guidance, this version is explicitly not blinded.}
}
\begin{document}

\maketitle

\begin{abstract}
We describe a unique environment in which undergraduate students from various STEM and social science disciplines are trained in data provenance and reproducible methods, and then apply that knowledge to real, conditionally accepted manuscripts and associated replication packages. We describe in detail the recruitment, training, and regular activities. While the activity is not part of a regular curriculum, the skills and knowledge taught through explicit training of reproducible methods and principles, and reinforced through repeated application in a real-life workflow, contribute to the education of these undergraduate students, and prepare them for post-graduation jobs and further studies.
\end{abstract}

\begin{quote}
\end{quote}

\section{Introduction}
\label{sec:intro}
The purpose of scientific publishing is the dissemination of robust research findings, exposing them to the scrutiny of peers. Key to this endeavor is documenting the provenance of those findings. Recent years have seen significant concerns expressed about the robustness of scientific results, commonly referred to as the ``replication crisis'' \citep{FanelliPNAS2018,gall_credibility_2017,hamermesh_viewpoint_2007,King1995-pg}. Various facets of the ``crisis'' have been explored (for just some of these, see \cite{olken_promises_2015,hamermesh_viewpoint_2007,StoddenPNAS2018,CamererScience2016}). Various approaches and solutions have been called for and proposed by the National Academies \citep{national_academies_of_sciences_engineering_and_medicine_reproducibility_2019}, committees of National Science Foundation \citep{Bollen2015}, and many scientists have called for greater transparency of research practices, and more assurance that published research is reproducible \citep{stodden_enhancing_2016,hoffler_replicationwiki_2017,Bell2013-rv,clemens_meaning_2017,coffman_proposal_2017}. Learned societies have a role to play within this discussion, as do journals.\footnote{For a collection of articles describing reproducibility in various disciplines, including economics \iftoggle{blind}{[REFERENCE blinded]}{\citep{vilhuber_reproducibility_2020}}, as well as an overview of the NASEM report cited above \citep{fineberg_highlights_2020}, see articles in the Harvard Data Science Review's \href{https://hdsr.mitpress.mit.edu/volume2issue4}{Fall 2020} issue.}


We should note here that the terms ``reproducible'' and ``replicable'' are not well-defined. Throughout this article, we will use them as defined in \citet{Bollen2015} and \citet{national_academies_of_sciences_engineering_and_medicine_reproducibility_2019}: (computational) reproducibility is ``obtaining consistent results using the same input data, computational steps, methods, and code, and conditions of analysis'' \citep[pg. 36]{national_academies_of_sciences_engineering_and_medicine_reproducibility_2019}. Replicability is achieved in this context through a relaxation of certain of the constraints implicit in the definition of reproducibility, for instance by collecting new data, implementing different methods or code, and then ``obtaining consistent results across studies aimed at answering the same scientific question [...] [obtaining] consistent results given the level of uncertainty inherent in the system under study'' (\textit{ibidem}). These definitions are broadly accepted in the social science and statistics community nowadays, but other communities may actually use these terms somewhat differently \citep[e.g.][]{heroux_editorial_2015}. We refer to ``replication packages'' as the collection of materials provided by authors to enable \textit{others} to replicate the results, but which should be ``reproducible'' themselves.


For empirical articles, the foundations on which they reside (data and its analysis) are external to the article and often to the journal they are published in. The data posting policies of many societies and journals, including the American Economic Association’s pre-2019 policy \citep{bernanke_editorial_2004}, were and are intended to create a minimal framework from which to replicate empirical findings. Historically, they have often failed. In several studies \citep{ChangAm.Econ.Rev.2017,CamererScience2016,HofflerAm.Econ.Rev.2017}, at least half of the replication packages associated with surveyed manuscripts failed to (fully) reproduce the results in the manuscript when re-run.

Increasingly, societies and journals have therefore switched to verifying and monitoring these policies \citep{JacobyInsideHigherEd2017,10.1257/pandp.109.718,editors_supporting_2021}. The \ac{AEA},  the  largest association of professional and academic economists in the world, with over 20,000 members located in 148 countries, has been at the forefront of such policies. It publishes 8 journals, including one of the top 5 journals in the discipline, in addition to several well-respected field journals. Concerns about the reliability and robustness of economic research have circulated in the AEA’s membership for more than 30 years \citep{Dewald1986-ni,mccullough_numerical_1999}. The policy to require that articles provide copies of their replication materials, first implemented in 2004 \citep{bernanke_editorial_2004}, was highly innovative at the time, but reflective of the membership's requests. Nevertheless, these early efforts improved the availability, but not necessarily the reproducibility of these replication packages. 
\iftoggle{blind}{[Sentence omitted for blinding.]}%
{In 2018, the Association appointed one of the authors of this article (Vilhuber) as the inaugural Data Editor \citep{10.1257/pandp.108.745}.} 
The Data Editor, in turn, started verifying, prior to final acceptance of an article, the computational reproducibility of the results displayed in the manuscript. The AEA’s endeavor is the largest in scale amongst the journals and societies conducting such verifications, having verified more than 1,000 articles since initiating such verifications two years ago.

The verification of replication packages, which  checks not only the computational reproducibility of the provided materials, but also verifies the documented provenance and completeness of such materials, is not a magical solution that will solve the ``reproducibility crisis.'' Replication packages may be reproducible, but wrong \citep[see the recent discussion surrounding][]{simonsohn_98_2021}. They do, however, reduce the cost for the scientific community to more easily find and assess such issues (finding the issue documented in \cite{simonsohn_98_2021} would have been much harder without a complete replication package) and can find other issues much earlier. For instance, the recent retraction in the \citet{journal_of_finance_retracted_2021} would have been detected prior to publication, not two years after publication. In the case of restricted-access data, pre-publication verification, when possible, may sometimes be the only opportunity to conduct such checks. Whether conducting reproducibility checks prior to publication is the most efficient or conducive exercise remains an open issue, including within the AEA's discussions on this topic. In this article, we describe the AEA's activities and how they contribute to this discussion.  


The \ac{AEA} began conducting comprehensive pre-publication reproducibility verification for conditionally-accepted manuscripts at its eight journals starting in 2019. These checks are conducted by the \ac{LDI} Replication Lab, which was set up by the current AEA Data Editor (Vilhuber), and which we describe in more detail in the next section. The Lab hires and trains undergraduate students who are primarily responsible for performing the required checks. 
%
The Lab's work with students is not integrated into any curriculum. Nevertheless, we will argue that students acquire some of the key data and computational skills described in \citet{national_academies_of_sciences_data_2018}. These abilities are the result of both the Lab's training and the observation of numerous completed but imperfect research projects. We also argue that students gain ``hands-on'' experience with some of the key dimensions of data-oriented science. The typical student will work on dozens of replication packages during their time at the Lab, with increasing autonomy along the way. These packages are of varying levels of complexity and completeness, and the students are required to assess their compliance with evolving and multi-faceted standards. This combination of taught and experiential learning provides the students with a strong foundation in data and computational management.  

The goal of this article is to describe the setting, the selection process for students, the actual workflow, and sketch out some of the observed outcomes. We hope to show that, while the setting may currently be fairly unique in its scale and position within the academic publication cycle, it is feasible to implement in a broader setting, and can meaningfully contribute to students' data science education. 

\section{Methods}
\subsection{Setting}
\label{sec:setting}

The \ac{AEA}'s \ac{DCAP} states that ``\textit{[i]t is the policy of the American Economic Association to publish papers only if the data and code used in the analysis are clearly and precisely documented and access to the data and code is non-exclusive to the authors}'' \citep{AmericanEconomicAssociation2019,10.1257/pandp.110.dcap}. 
To achieve the goal of improved (pervasive) computational reproducibility, 
the authors of conditionally accepted manuscripts are required to submit a replication package, consisting of all code used to process and analyze data, any data not available in an existing trusted repository, and a ``README'' describing data provenance and processing instructions.\footnote{The current requirements for authors are described in much detail at \href{https://perma.cc/FX2D-8FSW}{aeadataeditor.github.io/aea-de-guidance/}, but have varied over time. Confidential data are not part of the replication package, but must be described in the README, and are regularly made available to the Data Editor privately. As of February 2022, the README \textit{should} conform to \citet{vilhuber_lars_template_2020}, but this requirement, too, has varied over time.}  Each replication package is assessed  in terms of data provenance, clarity of the description, and computational reproducibility. These checks are conducted by the \ac{LDI} Replication Lab (henceforth ``the Lab''). 

The Lab was created by one of the authors of this article and AEA Data Editor (Vilhuber), based on earlier work starting in 2014, described in \citet{Kingi2018}.  
The Lab hires and trains undergraduate students who are primarily responsible for performing the required checks. These students are supervised by the Data Editor, the assistant Data Editor, and a graduate student. In a given calendar year, a total of about 50 undergraduate students work in the Lab with approximately 15-20 actively working on replications in a given week. Including the pilot phase, the Lab has hired over 100 \university{} undergraduate research assistants (RAs) to complete reproducibility verification checks.

While not a primary objective of the Lab's work for the AEA, the reliance on undergraduate students is intentional. The project provides undergraduate students with a unique opportunity to gain insights into the research and publication practices of hundreds of economists every year. Students are trained on reproducibility techniques, best practices, and communication skills, addressing several of the key dimensions highlighted in the \citet{national_academies_of_sciences_data_2018} report, namely ``data acumen'' and knowledge of some ``computational foundations,''  ``data management and curation,'' as well as ``workflow and reproducibility'' skills. Upon graduation, these students in turn take those skills into non-academic workplaces or graduate studies, thus seeding the next generation's improved practices. The Lab, therefore, attempts to target both ends of the academic publication pipeline: the ``outlet'', by improving replication packages of conditionally accepted publications, and the ``inlet'', by improving skills and attitudes of students at an early stage in their career. This article will focus on the latter.

The type of activity conducted at the Lab (systematic computational reproducibility checks within a publication workflow) is relatively novel. In particular, we are not aware of other institutions conducting such activities with undergraduate students. Other institutions that conduct reproducibility checks, such as the Odum Institute \citep{ChristianInt.J.Digit.Curation2018} and cascad \citep{perignon_certify_2019} primarily conduct such work with graduate students and professionals.

\subsection{Recruiting}
\label{sec:recruiting}

The Lab relies on students having some prior experience with statistical software, so as to be productive in short order. Despite the growth of open source software in the last decade, the software used in top economics journals is still overwhelmingly proprietary software (Stata, Matlab; see Figure 2 in \iftoggle{blind}{[REFERENCE blinded]}{\cite{vilhuber_reproducibility_2020}}). Furthermore, most usage by economists of said software is still strongly desktop-oriented. This is reflected in the replication packages received, and determines the required skill set of the students. However, the operating paradigms for most of these software packages remain quite similar. We have found that experience in any one of these packages is sufficient to allow students to follow instructions and conduct reproducibility verifications. Through prior experience, we have also found that knowledge of software more frequently used in computer science or engineering (for instance, Java, or C++) was not so useful, driving our current requirement of experience with what we know to be similar computing paradigms.  Of note, while we require some exposure to statistical software commonly used in the social sciences,  students are not expected to master it, nor to be proficient programmers.

These requirements are described in job postings, circulated among various undergraduate student experience coordinators on campus, and published to the campus-wide employment opportunity website and the \ac{LDI} website.\footnote{The posting as it appears on the LDI website is shown in Appendix \ref{sec:posting} and preserved at \url{https://perma.cc/G969-2HT4}.}
The posted job description provides applicants with information on requirements, wage rate, and maximum hours. As of January 2022, undergraduate RAs in the Lab are hired at the starting end of the Level II (out of four) classification wage rate, which, as of 12/31/2021, is \$13.45 per hour.\footnote{The full pay scale as it appears to students is reprinted in Appendix~\ref{sec:wagescale}, and preserved at \url{https://perma.cc/PTB5-98ZW}.} Hours are limited to 10 hours per week, which is less than the maximum number of hours allowed by \university{} policy.\footnote{Students can and do work more hours when classes are not in session - January and the summer months.}
Students are recruited from across campus, but most applicants are from the social sciences. Students apply with a cover letter and a CV, and are selected based primarily on observed and self-declared experience with one of the common statistical software packages used in the social sciences. The field they are majoring in plays no role, but is correlated with experience in these statistical software packages. Since most of the manuscripts stem from economics, most applicants have an interest, and often  a major in economics. No explicit years of study requirement is targeted. In practice, we have hired sophomores, juniors, and seniors, but not freshmen. In the most recent two rounds, we had 48 applications, and selected 21 for training. Of the 48 applications, 21 mentioned economics as one of their majors, 7 statistics, 9 computer or information science, and 21 none of those. Of the 21 students selected for training, 16 had an economics major, 5 had a statistics major, and 4 had a computer or information science major, with 4 having declared none of those majors.\footnote{Statistics generated from internal records as of 2021-09-15.}


\subsection{Training}
\label{sec:training}

    Although we announce the minimum requirements upon recruitment, the skill levels of the trainees are often heterogeneous, which could raise problems in conducting reproducibility assessments. Therefore, we train applicants before making final hiring decisions. The training is not remunerated, but is also not meant as a test. Rather, our intention with the training is to upskill all applicants. In practice, we retain over 90\% of trained applicants for at least one semester, and nearly all attrition in the past has been voluntary. 
    
    We currently provide training three times a year: before the fall and spring semesters, for students joining the Lab in that semester, as well as at the end of the spring semester for students joining the Lab as a summer job.  During this initial training, we provide instructions on all essential skills and knowledge necessary for the tasks. 
    The base training includes an overview lecture on the context of reproducibility concerns in economics, provides knowledge on reproducible practices, data provenance and data citations \citep{DataCitationSynthesisGroup2014}, includes  a presentation on the \citet{vilhuber_lars_template_2020} README,  basic instructions on command line and version control systems, and a detailed walkthrough of the assessment process, including how to prepare the final reproducibility report sent back to the authors.\footnote{The agenda for the January 2022 training can be viewed at \href{https://perma.cc/7WLN-NVJ4}{labordynamicsinstitute.github.io/replicability-training/}, most of the slides are available at \href{https://perma.cc/GP4P-YTTL?type=image}{labordynamicsinstitute.github.io/replicability-training-presentation}, and the textual content of training, which the presentation loosely follows, is available at \href{https://perma.cc/TP28-RDXL}{labordynamicsinstitute.github.io/replicability-training-curriculum/}.}
    It also reinforces computational skills, but does not train students on basic computational skills (since that was part of the selection criterion). Many of these topics are new to social science undergraduate students. 
    
    Depending on circumstances, in particular during the unusual 2020-2021 period, base training has been conducted as an intense in-person 8-hour training day, as a sequence of 2-4 hour training sessions spread over multiple days on Zoom, or even as an 8-hour long virtual training session. 
    The intense basic training is followed by a sequence of targeted test cases, interspersed with additional short lectures, reinforcing and deepening certain aspects (data provenance, debugging), as well as helping students acclimate to the Lab's task scheduling, reporting and workflow systems. Again, in adapting to changing circumstances over the last several semesters, these test cases and additional lectures have been concentrated into the three days immediately following the base training, or stretched out over the next three weeks, or only the following week. Students work on each test case on their own, following detailed step-by-step instructions, then interact with more experienced peers (undergraduate students already employed by the lab), and finally submit the report for each test case (as described in the Workflow) to the senior instructors and Lab leaders. Each case is discussed as a group before the next test case is presented and assigned, and students receive both generic feedback (commonly made errors or omissions) and individualized feedback.\footnote{The individualized feedback is generally prepared by one of the graduate assistants.}
    
    The first test case uses the dominant software package Stata, and introduces students to several small impediments to reproducibility. The case is a simple fake article with one table. The notion of ``add-on packages'' (libraries, etc., that need to be installed) is introduced. Stata, R, and many other statistical software rely on such packages, but they need to be specified for a replication package to be considered complete, since replicators may not have these installed. Authors of replication packages often neglect to identify such packages, and if not available, code will fail. The first test case uses one such package, but does not specify that it is needed. Students learn to recognize the type of error this generates, how to solve the problem, and how to document both the existence of the problem and the solution. Second, students are introduced to the idea of publicly accessible data that cannot be provided as part of a replication package. In this case, the package uses a dataset made available by ICPSR to researchers at member institutions. The terms of use specify that the data should not be redistributed. Students therefore need to download the data themselves. They learn to document any requirements (such as registration requirements, costs, application procedures, etc.), and to evaluate whether the description of the access conditions in a README is complete and sufficient. Finally, this test case is the first exposure to the structured replication template\footnote{\blindurl{https://github.com/AEADataEditor/replication-template}.}, the use of git and Markdown, and how to navigate the workflow process.
    
    The second test case uses the second-most frequent software in economics, MATLAB. Students are again provided with an article, except this time, it is a real article, with all its complexity and idiosyncracies. Students are now faced with identifying whether all code is provided for the tables, figures, and numbers in the article. It is a surprisingly frequent problem with replication packages that some significant part of the code is not provided. Sometimes, the missing code is used for appendices, sometimes it is data cleaning code, and sometimes, it is a key part of the overall replication package. In this case, both data cleaning code and code for some manuscript figures are missing. As for data,  all the data appear to be provided, and students are challenged to identify whether data provenance is sufficiently documented. Numerically, it is a ``small data'' case, so it is actually possible to compare the source data with the provided data. Finally, for many students, the use of MATLAB pushes most of them out of their comfort zone, and part of the training is to guide them to the similarities in user interfaces of statistical software, rather than focusing on the differences in the programming languages. The second test case is also the second time that the workflow is navigated, and the template used, instilling familiarity with the tools that will be used in the Lab.
    
    The third test case introduces students to more data provenance issues, while making it harder to verify that all the data are present. The case is a real article that uses confidential data that cannot be shared. Thus, students are challenged to identify whether all the data appear to be documented, by reading not just the README, but also the data section in the article. The article relies on Scandinavian register data and includes a thorough discussion of this data (a luxury not always present in other articles that students will later encounter). Students will assess whether data are properly cited, and write the report without ever running any code. 
    
    Both the second and third test cases rely on articles that were subsequently published.\footnote{The students work on publicly available pre-print versions of the articles, which do not include any of the recommendations made by the Data Editor and incorporated by the authors into the final version of record.} These cases were chosen because they give students practice with common issues they will face, while not being too computationally difficult or time consuming. The second test case emphasizes some of the technical skills needed to evaluate reproducibility (running code in an unfamiliar program, identifying parts of code in relation to tables/figures, etc), while the third test case helps students learn and practice how to identify data sources and verify packages for completeness without being able to run code. Students are shown how the final, published article and package differ from the earlier versions that they were provided with, and can see the impact that the reports can have on the clarity and reproducibility of scholarly publications. 
    
    As noted, each of the training test cases is accompanied by an initial presentation of the topic, significant independent work that relies on prepared documentation, a mentoring session with an experienced undergraduate replicator, and a debriefing session in which students provide their first impressions, and instructors provide both specific and generalized feedback. 
    
    At the end of the initial training, students join the regular Lab meetings, and are assigned real cases. However, we do generally take care to assign them cases that we believe to be easier to complete, slowly ramping up the complexity. We can do this because we have overlapping cohorts of students, with varying levels of experience, present in the Lab.

\subsection{Activities}
\label{sec:activities}

\begin{figure}
        \centering
        \includegraphics[width=0.9\textwidth]{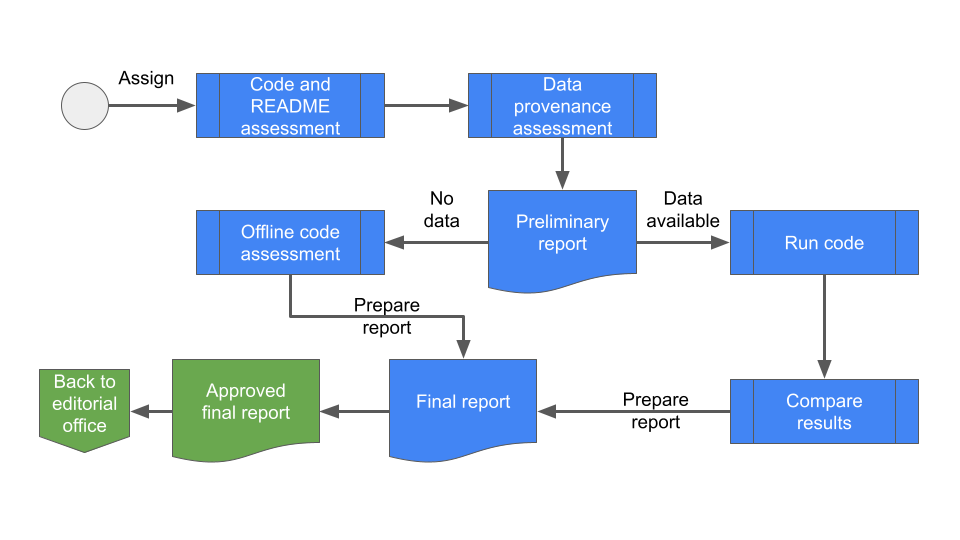}
        \caption{Simplified workflow for verification activities in the LDI Lab}
        \label{fig:workflow-simple}
    \end{figure}

    We now describe the generic workflow for the verification activities conducted in the Lab. A simplified workflow is shown in Figure~\ref{fig:workflow-simple}, a more detailed workflow with specific instructions for students is publicly available and regularly updated.%
    \footnote{\href{https://perma.cc/MK2R-J5PP}{labordynamicsinstitute.github.io/replicability-training-curriculum/aea-jira-workflow-a-guide.html}.} 
    The Lab receives a request to verify a replication package, generating a case number. Such replication packages generically consist of (or should consist of) computer code, possibly but rarely software, data that can be redistributed, and a document, generally referred to as the ``README,'' describing the provenance of all data, including data not made available as part of the package, and the process to reproduce the results in the paper. Materials are usually deposited by authors at the AEA Data and Code Repository\footnote{\url{https://www.openicpsr.org/openicpsr/aea}.}, though deposits made at other trusted repositories are also accepted (but are rare). The associated manuscript is also made available to the Lab.
    
    The case is assigned to the next available student. At this stage, no allowance is made for student skills or experience. The first part consists of an assessment of the README, the manuscript, and of the package contents.\footnote{We note that the AEA states that authors \textit{should} use the standardized README template \citep{vilhuber_lars_template_2020}. Many of the issues that the Lab encounters could be addressed by correct use of the README template.} Students assess whether data provenance is completely described, whether the manuscript and/or the README contain data citations \citep{DataCitationSynthesisGroup2014}, and whether there appears to be a complete archive of the program code (consistent with the README). Data provenance is assessed and verified, even when data are provided. When data are not provided, data provenance is even more important, as it must completely describe how the student can obtain the data. This can range from simple click-through downloads to complex application processes. The computer code and the README are assessed in terms of the software requirements. Finally, the computational resources are summarized, if reported in the README or deduced from other descriptions.
    
    This information is summarized in a preliminary report. One of the longest tasks is the data assessment task, since some manuscripts combine several dozen distinct data sources, and the quality of the description still ranges from excellent to poor. Once the preliminary report is finalized, it may be discussed with senior Lab members (team leads, graduate students, Data Editor): Does the assigned student have the necessary skills? Does the Lab have access to the software? Does the Lab have access to the computing resources (specific OS, high-performance computing)? Does the Lab have access to the data, or can it obtain access to the data without an unreasonable delay? Do the computations run in a reasonable amount of time, given all the resources available? Based on these assessments, the case may be reassigned to a student with the requisite experience, to a cooperating third party with access to data or computing resources, or, most frequently, remains with the original student.
    
    The student assigned to conduct the computational reproducibility check then downloads any necessary data, and follows instructions in the README to run the programs provided. This can be as simple as running a single ``main'' controller program, or as complex as running and baby-sitting dozens of programs, as well as conducting certain activities by hand. While much software can be automated, it is currently still very frequent to see certain automatable tasks conducted manually. Most often, this involves parsing computer output to generate tables (despite perfectly good code to do so programmatically) and generating maps using GIS software (despite perfectly good automated software to do so). It is the student's job to reproduce the manuscript's results following the authors' instructions as closely as possible, though this has some limits, in particular for very inefficient code or instructions.
    
    While all of the above steps appear to be very systematic, there are many subjective assessments. When are the manual steps too onerous? When is a deviation from stated computer runtimes unusual or too long, and action must be taken? The students are queried frequently about progress, and are required to discuss their active cases at twice-weekly meetings. In those meetings, all cases are discussed openly, and decisions are made. Those decisions do not necessarily come from the Data Editor, although he is present at most of these meetings. Students are encouraged to propose reasonable solutions, based on their experience, and increasingly make decisions autonomously. Input to such decisions may also come from team leads, or from graduate students. Students can also use a mailing list for peer-to-peer conversations and problems, and are encouraged to contact the student pre-approvers and the senior Lab members (Data Editor, assistant data editor, and graduate student) with more specific problems, which may be resolved in one-on-one meetings.
    
    Once the results have (or have not) been reproduced, the student completes a report. While based off of a template, the report has many free-form and narrative components, describing the steps the replicator undertook to achieve full computational reproducibility. Students are trained and mentored in how to convey steps in a concise but complete way, allowing the reader to understand fully what was done, and why it may not have worked. Students have access to a bank of frequently used ``canned responses''\footnote{The   ``canned responses'' can be found at \url{https://perma.cc/U8MR-WEEZ}. The latest version of the canned responses is available in the students' cloned template every time they initiate a new verification.} that provide constructive feedback and a checklist of requests. They are also mentored in providing an objective, positive, and dispassionate tone, as they are communicating with much more senior members of the profession. 
    
    The report is reviewed by more experienced students (``pre-approvers''),\footnote{We do not discuss the supplementary training for pre-approvers here.} and then reviewed again by the Data Editor before being sent to authors.\footnote{A sample report, suitably anonymized, can be viewed at \url{https://perma.cc/Z577-EEHG} and is included in Appendix~\ref{sec:sample-report}.} Students get feedback on how they performed on each case, and when improvements need to be made to the report or to the methods employed, they are provided both directly (privately) to the student, as well as in the form of generalized and anonymized counsel in the group meetings. 
    
    Most reports require that authors improve their package. In some cases, in particular when computational reproducibility was not achieved for key tables and figures, the replication package will go through another cycle of review after resubmission by the authors. When possible, such re-submissions are assessed by the same replicator. This is done both because it is efficient (the replicator will already have downloaded all the necessary data, often many gigabytes, into their personal replication workspace) and because it provides validating feedback to the replicator, showing them that authors are (generally) quite responsive to constructive comments. Students then write an abbreviated revision report, which goes through the same review cycle as the full initial report. 
    
    The cycle described above gets repeated for every new case assigned to the Lab. The typical case will be in the student's hands for about 10-14 days, though they may sometimes work on 2 or more cases simultaneously, as they wait for code to run or data access requests to be authorized, and some cases may take significantly longer. 
    
    To give context, we refer the reader to the AEA Data Editor's annual reports \citep{10.1257/pandp.109.718,VilhuberAEAPap.Proc.2020,vilhuberReportAEAData2021,aeareport2022}. In the most recent year, the LDI Lab received 529 requests for 415 manuscripts \citep{aeareport2022}. The vast majority of manuscripts go through a single round of reviews, typically with minor changes requested \citep[Tables~2 and 3]{aeareport2022}. The median time to full acceptance is between 4 and 6 weeks (reported in various annual reports by the AEA journals).\footnote{As an aside, when replication packages are not unconditionally accepted, changes that are requested almost always include both corrections to data provenance descriptions (incomplete access description, absent data citations) and minor, fixable corrections to code (directories not created, requirements not fully described). }



\section{Discussion}
\label{sec:discussion}

The recruitment, training, and regular activities have been refined in an iterative process since early 2018. After each training, we have reviewed the effectiveness of our methods, surveyed trainees about their perception of the training, and incorporated improvements into the next round. For instance, the peer-driven tutorial as part of the test cases was based on trainee feedback, and has proven both popular and effective.

\subsection{Student development opportunities within the Lab}

    As students accumulate cases, we have observed fairly rapid gains in maturity and autonomy. 
    The best students may be promoted to team leads, where they run the shorter check-in meetings mostly autonomously, are expected to provide a first level of support to their team members, and may be asked to be ``pre-approvers.'' Students contribute regularly to overall improvements in processes and procedures, and are encouraged to contribute to a Wiki, with the intent of providing a knowledge base that is driven by the students for the students. In certain circumstances, students may provide short (verbal or written) tutorials or presentations at group meetings when they have solved a particularly thorny problem, for which the solution is of potential utility to all the students.  
    
\subsection{Student-faculty interaction at the Lab}

    Two types of student-faculty interaction occur during the Lab activities in which the students participate. First, students meet twice a week with the faculty supervisor (the Data Editor), to discuss individual cases. These meetings are group meetings, and take as long as necessary to guide the student onto the right path. Where appropriate, one-on-one meetings with the faculty supervisor or graduate students are also scheduled, to solve thornier problems. 
    
    The second type of interaction is more indirect. All reports written by students, once approved, are read by manuscript authors, and may be read by journal editors. Students also read the authors’ responses to any requested changes. Authors and editors are most often academic faculty. Students do not, however, interact directly with authors and editors - in fact, this is explicitly forbidden by the rules of behavior of the Lab. Thus, this indirect interaction with academic and private-sector faculty is always mediated via structured written communication.

\subsection{Relationship to other curricular activities}

The activities described here are not offered as a regular for-credit course. Rather, they are offered as an on-campus, research-oriented job. As such, they do not neatly fit into a curricular development or plan. However, our impression is that students gain valuable experience through participation in the Lab's activities. This (admittedly subjective and potentially biased) opinion relies on a few observations. For one, almost all students stay with the Lab until graduation, and sometimes even beyond graduation. They also tend to continue at the Lab over the summer, despite having other summer jobs or internships (the hours are adjusted to accommodate such constraints). Such implicit ``job satisfaction'' indicates that students believe that the experience is valuable, despite having access to higher-paying jobs during summers and post-graduation. 

Second, while most students come to us with some prior exposure to programming and data analysis, they largely need training on both the conceptual and technical skills necessary to assess reproducibility. This suggests that they are not currently acquiring these skills elsewhere in their coursework. While the various curricula on campus do offer some of these elements explicitly, and other courses likely embed these techniques within their more discipline-specific topics, there does seem to be a need to more generally expose undergraduate students to these techniques. The Lab's training, while not designed to provide training in such techniques, seems to complement other curricular offerings. For example, while many students have read academic papers in other courses, they rarely have the skill set upon joining the lab to identify and summarize the data sources used in the paper. This skill is something that we train students on in order to evaluate the existence or completeness of data citations.

\subsection{Longer term outcomes}

We have not previously conducted a formal evaluation of the takeaways and experiences undergraduates have had after working for the Lab, and we do not systematically conduct exit interviews with undergraduates at the end of their job tenure. We have, however, conducted informal conversations with the goal of a formative evaluation of initial training and later activities. A few graduates have reported   that they have ``learned a lot about reproducibility'' in ways that ``really helped me as a research assistant'' (2021 economics graduate working for a large national economic consulting firm), and that they received ``overwhelmingly positive feedback on my documentation method in code reviews, which is all thanks to my time with \iftoggle{blind}{[BLINDED]}{LDI}'' (2020 sociology graduate working for a non-profit research organization). A more formal survey of former Lab members is currently in the field, and we plan to report on outcomes in the future. 

One of the difficulties of empirically measuring the effect of the explicit initial training and the implicit on-the-job training provided is the noisiness of most empirical measures. Students are not required to actively program, but they will learn programming techniques. No student will have been exposed to data citation principles prior to joining the Lab, and yet all will have learned and applied such principles by the end of the initial training, and refined it over time. The students' efficiency at conducting computational reproducibility invariably will increase, but allocation of papers is not random, and the difficulty of papers varies so widely that any objective measure of time or effort will likely be too noisy to be useful. We continue considering ways to measure this in the future.

\subsection{Generalization of Lab activities}

When initially planning how to translate the AEA's verification activities into a feasible operation, the reliance on undergraduate students was intentional, for two reasons. For one, we believed that with proper training, undergraduates would provide a more cost effective verification of the basic computational reproducibility of the packages we received than with a rotating cadre of graduate students. This has largely been born out, though we do not provide evidence here on the financial underpinnings of the operation. More importantly, however, we intentionally forced ourselves to develop a training program for those undergraduates, believing that it would have utility for other universities, disciplines, journals, and in other circumstances. 



If an instructor wanted to directly integrate these activities into a curriculum, it is likely best to implement them as a formal course. This course should include the training that research assistants currently receive as well as training for how to actively create reproducible packages on their own. The initial training alone accounts for about 14 hours of classroom instruction, plus significant ``homework'' time. With a straightforward and pedagogically valuable expansion of some of the themes that get short shrift in the current training because they are taught and learned in the first ``real cases'' (git, version control, objective communication skills), a course based on these materials would easily cover 21 hours of classroom training. Test cases could be expanded to have higher and more varied computational requirements, and can easily be based on real replication packages, suitably modified to highlight specific learning objectives. Student assessments could be based on recognition of required components of reproducible packages, followed by written reports on actual replication packages, and concluding with the students creating their own replication package, based on either a research project from another course or a novel one as part of this course. If using a research project from another course, then this replication course could be paired with, for example, an upper level course with a term paper component. 

While it may seem attractive to embed some of the minor activities into an existing course, our experience with the particular student population from which we recruited leads us to conclude that the necessary start-up training is intensive enough to require a standalone course. We note, however, that almost none of our recruits have taken more explicitly data-science oriented courses, as far as we can tell. Our suggestions are thus likely not representative when more technically advanced students are included in the population of interest.

We do, however, believe that what we observe is not specific to economics, and could be easily implemented in other (related) disciplines. Discussions with colleagues in sociology or policy suggests that the basic context and student skill set may be quite similar there. Amongst our recruits have been engineering, bio-statistics, and sociology students, and they have performed just as well as the students with an economics major. 

One final thought, though. Implementing the activities outlined here as part of a course is unlikely to then also meet the needs of a journal for reliable and timely verification service. Most journals would have verification needs at monthly or even weekly frequencies throughout the year, including when classes are not in session. A course would not even produce a continuous stream of verification reports throughout the semester. A course might, however, serve as a feeder to a campus-wide verification service, similar to statistical consulting services with student workers. 

\section{Acknowledgements}
\label{sec:acks}
We thank 
\iftoggle{blind}{[BLINDED {\it former graduate students}]}%
{Hautahi Kingi, Sylverie Herbert, and Flavio Stanchi}, who contributed to the earliest pilots of this effort. We thank the students who have worked for the Lab, and who have helped improve both the reproducibility of economics articles, as well as the training and workflow for later participants.

\section{Declaration of Interests Statement}
\label{sec:doi}
\iftoggle{blind}{

There are DoI in the non-blinded version.
}{%
LV is the Data Editor of the American Economic Association, a paid position. All activities described herein are funded by the American Economic Association. HHS, MW, and DNW were graduate student assistants at various points during the time period described in the article, and MD is the assistant data editor, all hired by LV and paid through the AEA's contract with Cornell University. 
}


\printbibliography

@article{aeareport2022,
  title = {Report by the {{AEA Data Editor}}},
  author = {Vilhuber, Lars},
  date = {2022-05},
  journaltitle = {AEA Papers and Proceedings},
  volume = {112},
  doi = {10.1257/pandp.112.XXXXX},
}

@article{10.1257/pandp.109.718,
  title = {Report by the {{AEA Data Editor}}},
  author = {Vilhuber, Lars},
  date = {2019-05},
  journaltitle = {AEA Papers and Proceedings},
  volume = {109},
  pages = {718--29},
  issn = {2574-0768, 2574-0776},
  doi = {10.1257/pandp.109.718},
  url = {http://www.aeaweb.org/articles?id=10.1257/pandp.109.718},
  urldate = {2019-09-21},
  langid = {english}
}

@article{VilhuberAEAPap.Proc.2020,
  title = {Report by the {{AEA Data Editor}}},
  author = {Vilhuber, Lars and Turitto, James and Welch, Keesler},
  date = {2020-05},
  journaltitle = {AEA Papers and Proceedings},
  volume = {110},
  pages = {764--75},
  issn = {2574-0768, 2574-0776},
  doi = {10.1257/pandp.110.764},
  langid = {english}
}

@article{vilhuberReportAEAData2021,
  title = {Report by the {{AEA Data Editor}}},
  author = {Vilhuber, Lars},
  date = {2021-05-01},
  journaltitle = {AEA Papers and Proceedings},
  shortjournal = {AEA Papers and Proceedings},
  volume = {111},
  pages = {808--817},
  issn = {2574-0768, 2574-0776},
  doi = {10.1257/pandp.111.808},
  url = {https://pubs.aeaweb.org/doi/10.1257/pandp.111.808},
  urldate = {2021-05-20},
  langid = {english}
}

@article{gall_credibility_2017,
	title = {The credibility crisis in research: {Can} economics tools help?},
	volume = {15},
	issn = {1545-7885},
	shorttitle = {The credibility crisis in research},
	url = {http://journals.plos.org/plosbiology/article?id=10.1371/journal.pbio.2001846},
	doi = {10.1371/journal.pbio.2001846},
	number = {4},
	urldate = {2017-12-22},
	journal = {PLOS Biology},
	author = {Gall, Thomas and Ioannidis, John P. A. and Maniadis, Zacharias},
	month = apr,
	year = {2017},
	pages = {e2001846},
}

@article{FanelliPNAS2018,
	title = {Opinion: {Is} science really facing a reproducibility crisis, and do we need it to?},
	volume = {115},
	copyright = {© 2018 . Published under the PNAS license.},
	issn = {0027-8424, 1091-6490},
	shorttitle = {Opinion},
	url = {http://www.pnas.org/content/115/11/2628},
	doi = {10.1073/pnas.1708272114},
	language = {en},
	number = {11},
	urldate = {2018-05-31},
	journal = {Proceedings of the National Academy of Sciences},
	author = {Fanelli, Daniele},
	month = mar,
	year = {2018},
	pmid = {29531051},
	pages = {2628--2631},
}

@book{national_academies_of_sciences_data_2018,
	title = {Data {Science} for {Undergraduates}: {Opportunities} and {Options}},
	isbn = {978-0-309-47559-4},
	shorttitle = {Data {Science} for {Undergraduates}},
	url = {https://www.nap.edu/catalog/25104/data-science-for-undergraduates-opportunities-and-options},
	language = {en},
	urldate = {2021-07-05},
	author = {National Academies of Sciences, Engineering},
	month = may,
	year = {2018},
	doi = {10.17226/25104},
}

@techreport{Kingi2018,
	address = {Berkeley, CA},
	type = {Presentation},
	title = {The {Reproducibility} of {Economics} {Research}:  {A} {Case} {Study}},
	copyright = {All rights reserved},
	url = {https://osf.io/srg57/},
	author = {Kingi, Hautahi and Stanchi, Flavio and Vilhuber, Lars and Herbert, Sylverie},
	year = {2018},
}

@article{heroux_editorial_2015,
	title = {Editorial: {ACM} {TOMS} {Replicated} {Computational} {Results} {Initiative}},
	volume = {41},
	issn = {0098-3500, 1557-7295},
	shorttitle = {Editorial},
	url = {https://dl.acm.org/doi/10.1145/2743015},
	doi = {10.1145/2743015},
	language = {en},
	number = {3},
	urldate = {2022-02-13},
	journal = {ACM Transactions on Mathematical Software},
	author = {Heroux, Michael A.},
	month = jun,
	year = {2015},
	pages = {1--5},
}

@techreport{DataCitationSynthesisGroup2014,
	title = {Joint {Declaration} of {Data} {Citation} {Principles}},
	institution = {Force11},
	author = {{Data Citation Synthesis Group} and Martone, Maryann},
	year = {2014},
	doi = {10.25490/a97f-egyk},
}

@article{perignon_certify_2019,
	title = {Certify reproducibility with confidential data},
	volume = {365},
	copyright = {Copyright © 2019, American Association for the Advancement of Science. http://www.sciencemag.org/about/science-licenses-journal-article-reuseThis is an article distributed under the terms of the Science Journals Default License.},
	issn = {0036-8075, 1095-9203},
	url = {https://science.sciencemag.org/content/365/6449/127},
	doi = {10.1126/science.aaw2825},
	language = {en},
	number = {6449},
	urldate = {2019-09-22},
	journal = {Science},
	author = {Pérignon, Christophe and Gadouche, Kamel and Hurlin, Christophe and Silberman, Roxane and Debonnel, Eric},
	month = jul,
	year = {2019},
	pmid = {31296759},
	pages = {127--128},
}

@misc{AmericanEconomicAssociation2019,
	title = {Data and {Code} {Availability} {Policy}},
	url = {https://www.aeaweb.org/journals/policies/data-code},
	urldate = {2019-11-19},
	author = {{American Economic Association}},
	month = jul,
	year = {2019},
}

@article{10.1257/pandp.110.dcap,
	title = {Data and code availability policy},
	volume = {110},
	doi = {10.1257/pandp.110.776},
	journal = {AEA Papers and Proceedings},
	author = {{American Economic Association}},
	month = may,
	year = {2020},
	pages = {776--78},
}

@article{journal_of_finance_retracted_2021,
	title = {Retracted: {Risk} {Management} in {Financial} {Institutions}},
	volume = {n/a},
	issn = {1540-6261},
	shorttitle = {Retracted},
	url = {https://onlinelibrary.wiley.com/doi/abs/10.1111/jofi.13064},
	doi = {10.1111/jofi.13064},
	language = {en},
	number = {n/a},
	urldate = {2021-08-22},
	journal = {The Journal of Finance},
	author = {{Journal of Finance}},
	year = {2021},
	note = {\_eprint: https://onlinelibrary.wiley.com/doi/pdf/10.1111/jofi.13064},
}

@misc{simonsohn_98_2021,
	title = {[98] {Evidence} of {Fraud} in an {Influential} {Field} {Experiment} {About} {Dishonesty}},
	url = {https://datacolada.org/98},
	language = {en-US},
	urldate = {2021-08-22},
	journal = {Data Colada},
	author = {Simonsohn, Uri and Nelson, Leif and Simmons, Joe and {Anonymous}},
	month = aug,
	year = {2021},
}

@article{10.1257/pandp.108.745,
	title = {Report of the {Search} {Committee} to {Appoint} a {Data} {Editor} for the {AEA}},
	volume = {108},
	issn = {2574-0768},
	url = {http://www.aeaweb.org/articles?id=10.1257/pandp.108.745},
	doi = {10.1257/pandp.108.745},
	language = {en},
	urldate = {2018-07-22},
	journal = {AEA Papers and Proceedings},
	author = {Duflo, Esther and Hoynes, Hilary},
	year = {2018},
	pages = {745},
}

@article{mccullough_numerical_1999,
	title = {The {Numerical} {Reliability} of {Econometric} {Software}},
	volume = {37},
	issn = {0022-0515},
	url = {https://www.aeaweb.org/articles?id=10.1257/jel.37.2.633},
	doi = {10.1257/jel.37.2.633},
	language = {en},
	number = {2},
	urldate = {2019-01-28},
	journal = {Journal of Economic Literature},
	author = {McCullough, B. D. and Vinod, H. D.},
	month = jun,
	year = {1999},
	pages = {633--665},
}

@article{Dewald1986-ni,
	title = {Replication in {Empirical} {Economics}: {The} {Journal} of {Money}, {Credit} and {Banking} {Project}},
	volume = {76},
	issn = {0002-8282},
	url = {https://www.jstor.org/stable/1806061},
	number = {4},
	journal = {The American Economic Review},
	author = {Dewald, William G and Thursby, Jerry G and Anderson, Richard G},
	year = {1986},
	pages = {587--603},
}

@article{editors_supporting_2021,
	title = {Supporting computational reproducibility through code review},
	volume = {5},
	issn = {2397-3374},
	url = {https://www.nature.com/articles/s41562-021-01190-w},
	doi = {10.1038/s41562-021-01190-w},
	language = {en},
	number = {8},
	urldate = {2021-08-21},
	journal = {Nature Human Behaviour},
	author = {{Editors}},
	month = aug,
	year = {2021},
	pages = {965--966},
}

@article{ChristianInt.J.Digit.Curation2018,
	title = {Operationalizing the {Replication} {Standard}: {A} {Case} {Study} of the {Data} {Curation} and {Verification} {Workflow} for {Scholarly} {Journals}},
	volume = {13},
	shorttitle = {Operationalizing the {Replication} {Standard}},
	url = {https://osf.io/preprints/socarxiv/cfdba/},
	doi = {10.2218/ijdc.v13i1.555},
	number = {1},
	urldate = {2018-06-04},
	journal = {International Journal of Digital Curation},
	author = {Christian, Thu-Mai and Lafferty-Hess, Sophia and Jacoby, William and Carsey, Thomas},
	year = {2018},
}

@misc{JacobyInsideHigherEd2017,
	title = {Should {Journals} {Be} {Responsible} for {Reproducibility}?},
	shorttitle = {Should {Journals} {Be} {Responsible} for {Reproducibility}?},
	url = {https://www.insidehighered.com/blogs/rethinking-research/should-journals-be-responsible-reproducibility},
	language = {en},
	urldate = {2018-07-22},
	journal = {Inside Higher Ed},
	author = {Jacoby, William G. and Lafferty-Hess, Sophia and Christian, Thu-Mai},
	month = jul,
	year = {2017},
}

@article{ChangAm.Econ.Rev.2017,
	title = {A {Preanalysis} {Plan} to {Replicate} {Sixty} {Economics} {Research} {Papers} {That} {Worked} {Half} of the {Time}},
	volume = {107},
	issn = {0002-8282},
	url = {https://www.aeaweb.org/articles?id=10.1257/aer.p20171034},
	doi = {10.1257/aer.p20171034},
	language = {en},
	number = {5},
	urldate = {2018-05-22},
	journal = {American Economic Review},
	author = {Chang, Andrew C. and Li, Phillip},
	month = may,
	year = {2017},
	pages = {60--64},
}

@article{bernanke_editorial_2004,
	title = {Editorial {Statement}},
	volume = {94},
	issn = {0002-8282},
	url = {https://www.jstor.org/stable/3592790},
	number = {1},
	urldate = {2020-09-01},
	journal = {The American Economic Review},
	author = {Bernanke, Ben S.},
	year = {2004},
	note = {Publisher: American Economic Association},
	pages = {404--404},
}

@article{King1995-pg,
	title = {Replication, {Replication}},
	volume = {28},
	issn = {1049-0965},
	number = {3},
	journal = {PS, political science \& politics},
	author = {King, Gary},
	month = sep,
	year = {1995},
	pages = {443--499},
}

@article{fineberg_highlights_2020,
	title = {Highlights of the {US} {National} {Academies} {Report} on “{Reproducibility} and {Replicability} in {Science}”},
	volume = {2},
	issn = {,},
	url = {https://hdsr.mitpress.mit.edu/pub/6an6ppum/release/4},
	doi = {10.1162/99608f92.cb310198},
	language = {en},
	number = {4},
	urldate = {2021-08-22},
	journal = {Harvard Data Science Review},
	author = {Fineberg, Harvey and Stodden, Victoria and Meng, Xiao-Li},
	month = oct,
	year = {2020},
	note = {Publisher: PubPub},
}

@article{vilhuber_lars_template_2020,
	title = {A template {README} for social science replication packages},
	copyright = {Creative Commons Attribution Non Commercial 4.0 International, Open Access},
	url = {https://zenodo.org/record/4319999},
	doi = {10.5281/ZENODO.4319999},
	language = {en},
	urldate = {2021-04-01},
	author = {Vilhuber, Lars and Connolly, Marie and Koren, Miklós and Llull, Joan and Morrow, Peter},
	month = dec,
	year = {2020},
	note = {Publisher: Zenodo
Version Number: v1.0.0},
}

@article{vilhuber_reproducibility_2020,
	title = {Reproducibility and {Replicability} in {Economics}},
	volume = {2},
	copyright = {All rights reserved},
	url = {https://hdsr.mitpress.mit.edu/pub/fgpmpj1l},
	doi = {10.1162/99608f92.4f6b9e67},
	number = {4},
	journal = {Harvard Data Science Review},
	author = {Vilhuber, Lars},
	year = {2020},
}

@article{HofflerAm.Econ.Rev.2017,
	title = {Replication and {Economics} {Journal} {Policies}},
	volume = {107},
	issn = {0002-8282},
	url = {https://www.aeaweb.org/articles?id=10.1257/aer.p20171032},
	doi = {10.1257/aer.p20171032},
	language = {en},
	number = {5},
	urldate = {2018-05-22},
	journal = {American Economic Review},
	author = {Höffler, Jan H.},
	month = may,
	year = {2017},
	pages = {52--55},
}

@article{hoffler_replicationwiki_2017,
	title = {{ReplicationWiki}: {Improving} {Transparency} in {Social} {Sciences} {Research}},
	volume = {23},
	issn = {1082-9873},
	shorttitle = {{ReplicationWiki}},
	url = {http://www.dlib.org/dlib/march17/hoeffler/03hoeffler.html},
	doi = {10.1045/march2017-hoeffler},
	language = {en},
	number = {3/4},
	urldate = {2018-07-17},
	journal = {D-Lib Magazine},
	author = {Höffler, Jan H.},
	month = mar,
	year = {2017},
}

@article{coffman_proposal_2017,
	title = {A {Proposal} to {Organize} and {Promote} {Replications}},
	volume = {107},
	issn = {0002-8282},
	url = {https://www.aeaweb.org/articles?id=10.1257/aer.p20171122},
	doi = {10.1257/aer.p20171122},
	language = {en},
	number = {5},
	urldate = {2018-05-22},
	journal = {American Economic Review},
	author = {Coffman, Lucas C. and Niederle, Muriel and Wilson, Alistair J.},
	month = may,
	year = {2017},
	pages = {41--45},
}

@article{clemens_meaning_2017,
	title = {The {Meaning} of {Failed} {Replications}: {A} {Review} and {Proposal}},
	volume = {31},
	copyright = {© 2015 John Wiley \& Sons Ltd},
	issn = {1467-6419},
	shorttitle = {The {Meaning} of {Failed} {Replications}},
	url = {https://onlinelibrary.wiley.com/doi/abs/10.1111/joes.12139},
	doi = {10.1111/joes.12139},
	language = {en},
	number = {1},
	urldate = {2018-05-22},
	journal = {Journal of Economic Surveys},
	author = {Clemens, Michael A.},
	month = feb,
	year = {2017},
	pages = {326--342},
}

@article{Bell2013-rv,
	title = {How to {Persuade} {Journals} to {Accept} {Your} {Replication} {Paper}},
	url = {https://politicalsciencereplication.wordpress.com/2013/09/11/guest-blog-how-to-persuade-journals-to-accept-your-replication-paper/},
	author = {Bell, Mark and Miller, Nicholas},
	year = {2013},
}

@article{stodden_enhancing_2016,
	title = {Enhancing reproducibility for computational methods},
	volume = {354},
	copyright = {Copyright © 2016, American Association for the Advancement of Science},
	issn = {0036-8075, 1095-9203},
	url = {http://science.sciencemag.org/content/354/6317/1240},
	doi = {10.1126/science.aah6168},
	language = {en},
	number = {6317},
	urldate = {2018-05-20},
	journal = {Science},
	author = {Stodden, Victoria and McNutt, Marcia and Bailey, David H. and Deelman, Ewa and Gil, Yolanda and Hanson, Brooks and Heroux, Michael A. and Ioannidis, John P. A. and Taufer, Michela},
	month = dec,
	year = {2016},
	pmid = {27940837},
	pages = {1240--1241},
}

@book{national_academies_of_sciences_engineering_and_medicine_reproducibility_2019,
	address = {Washington, D.C.},
	title = {Reproducibility and {Replicability} in {Science}},
	isbn = {978-0-309-48616-3},
	url = {https://doi.org/10.17226/25303},
	urldate = {2019-09-21},
	publisher = {National Academies Press},
	author = {{National Academies of Sciences, Engineering, and Medicine}},
	year = {2019},
	doi = {10.17226/25303},
}

@techreport{Bollen2015,
	type = {Report of the {Subcommittee} on {Replicability} in {Science} {Advisory} {Committee} to the {National} {Science} {Foundation} {Directorate} for {Social}, {Behavioral}, and {Economic} {Sciences}},
	title = {Social, {Behavioral}, and {Economic} {Sciences} {Perspectives} on {Robust} and {Reliable} {Science}},
	url = {https://www.nsf.gov/sbe/AC_Materials/SBE_Robust_and_Reliable_Research_Report.pdf},
	urldate = {2018-05-20},
	institution = {National Science Foundation},
	author = {Bollen, Kenneth and Cacioppo, John T. and Kaplan, Robert M. and Krosnick, Jon A. and Olds, James L.},
	year = {2015},
}

@article{CamererScience2016,
	title = {Evaluating replicability of laboratory experiments in economics},
	copyright = {Copyright © 2016, American Association for the Advancement of Science},
	issn = {0036-8075, 1095-9203},
	url = {http://science.sciencemag.org/content/early/2016/03/02/science.aaf0918},
	doi = {10.1126/science.aaf0918},
	language = {en},
	urldate = {2018-06-04},
	journal = {Science},
	author = {Camerer, Colin F. and Dreber, Anna and Forsell, Eskil and Ho, Teck-Hua and Huber, Jürgen and Johannesson, Magnus and Kirchler, Michael and Almenberg, Johan and Altmejd, Adam and Chan, Taizan and Heikensten, Emma and Holzmeister, Felix and Imai, Taisuke and Isaksson, Siri and Nave, Gideon and Pfeiffer, Thomas and Razen, Michael and Wu, Hang},
	month = mar,
	year = {2016},
	pmid = {26940865},
	pages = {aaf0918},
}

@article{StoddenPNAS2018,
	title = {An empirical analysis of journal policy effectiveness for computational reproducibility},
	copyright = {© 2018 . Published under the PNAS license.},
	issn = {0027-8424, 1091-6490},
	url = {http://www.pnas.org/content/early/2018/03/08/1708290115},
	doi = {10.1073/pnas.1708290115},
	language = {en},
	urldate = {2018-05-20},
	journal = {Proceedings of the National Academy of Sciences},
	author = {Stodden, Victoria and Seiler, Jennifer and Ma, Zhaokun},
	month = mar,
	year = {2018},
	pmid = {29531050},
	pages = {201708290},
}

@article{olken_promises_2015,
	title = {Promises and {Perils} of {Pre}-analysis {Plans}},
	volume = {29},
	issn = {0895-3309},
	url = {https://www.aeaweb.org/articles?id=10.1257/jep.29.3.61},
	doi = {10.1257/jep.29.3.61},
	language = {en},
	number = {3},
	urldate = {2020-06-15},
	journal = {Journal of Economic Perspectives},
	author = {Olken, Benjamin A.},
	month = sep,
	year = {2015},
	note = {tex.ids= olken2015},
	pages = {61--80},
}

@article{hamermesh_viewpoint_2007,
	title = {Viewpoint: {Replication} in economics},
	volume = {40},
	issn = {1540-5982},
	shorttitle = {Viewpoint},
	url = {https://onlinelibrary.wiley.com/doi/abs/10.1111/j.1365-2966.2007.00428.x},
	doi = {10.1111/j.1365-2966.2007.00428.x},
	language = {en},
	number = {3},
	urldate = {2018-05-21},
	journal = {Canadian Journal of Economics},
	author = {Hamermesh, Daniel S.},
	month = aug,
	year = {2007},
	pages = {715--733},
}
\newpage
\appendix

\section{Job posting on institution webpage}
\label{sec:posting}

The following image of the job posting was captured on February 6, 2022. It is preserved at \url{https://perma.cc/G969-2HT4}.

\includegraphics[page=1,width=0.9\textwidth]{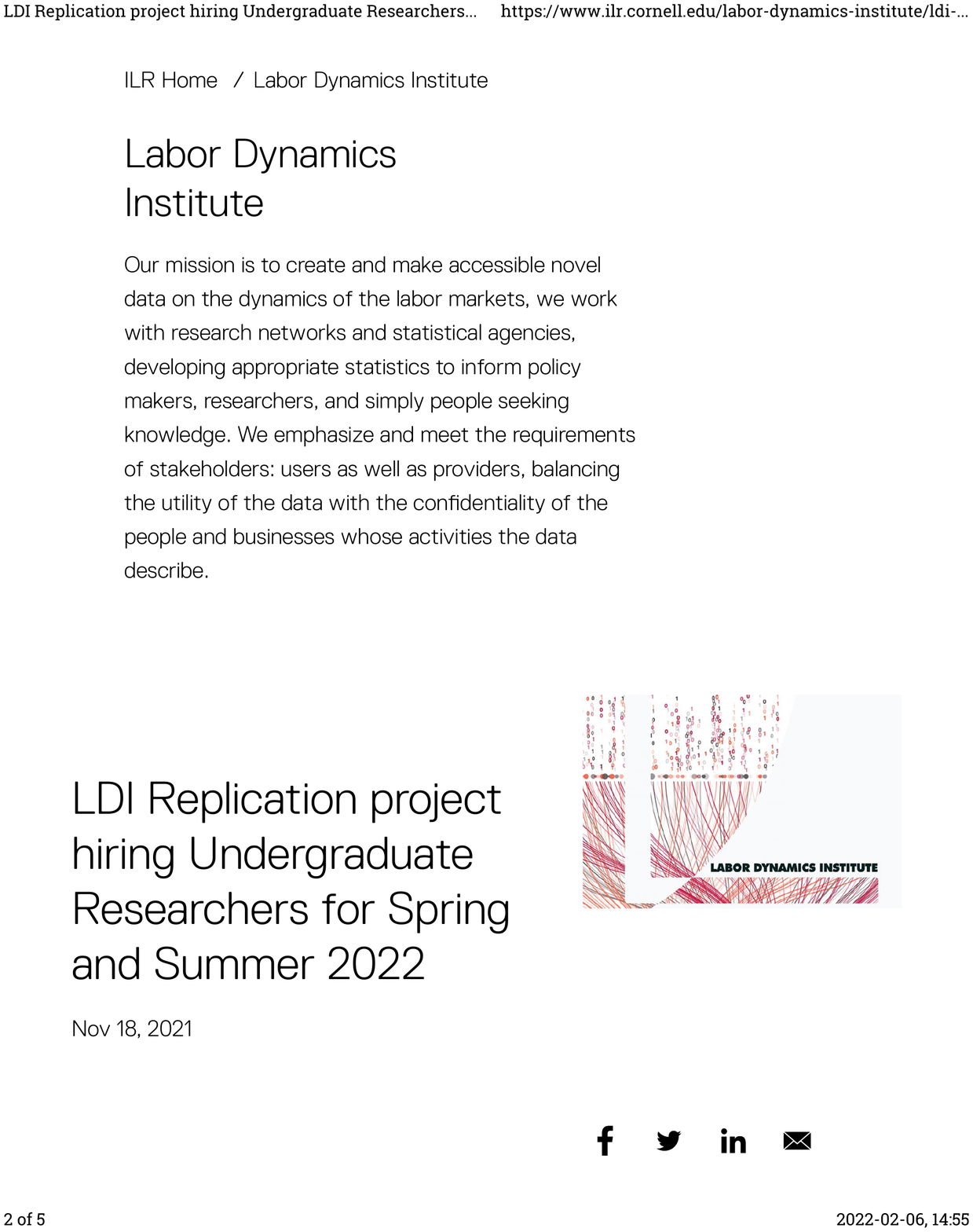}

\includegraphics[page=2,width=0.9\textwidth]{appendix-job-posting-2022.pdf}

\newpage

\section{Wage scale range at the university level}
\label{sec:wagescale}

The following image of the job posting was captured on February 6, 2022. It is preserved at \url{https://perma.cc/PTB5-98ZW}.

\includegraphics[page=1,width=0.9\textwidth]{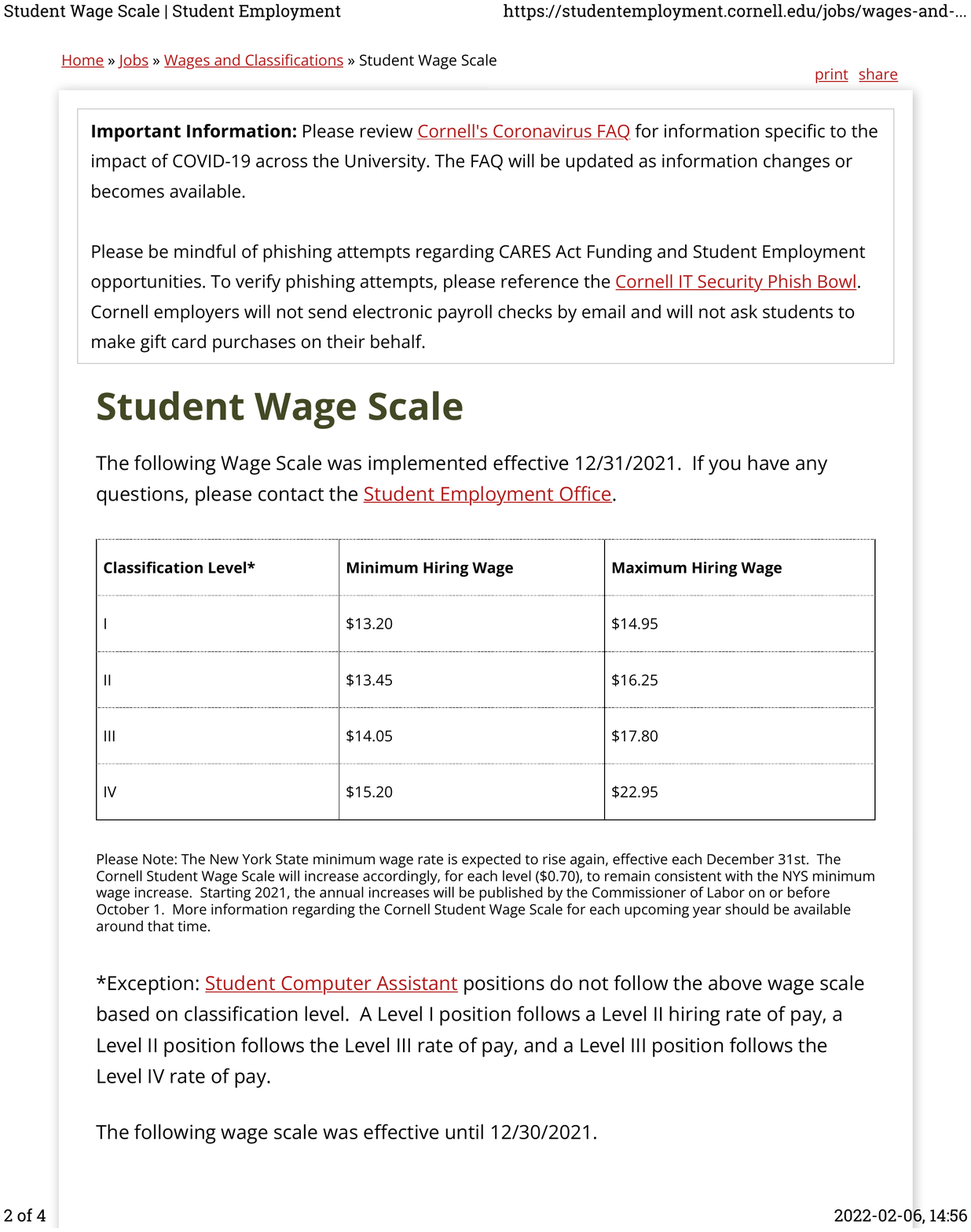}

\includegraphics[page=2,width=0.9\textwidth]{appendix-cornell-Student-Wage-Scale.pdf}

\section{Sample reproducibility report sent to authors}
\label{sec:sample-report}

The following report has been anonymized, but reflects the typical report sent back to authors. Length depends on number of data sources (each data source is listed) and number of programs (each is listed, together with reproducibility status). 

\includegraphics[page=1,width=0.9\textwidth]{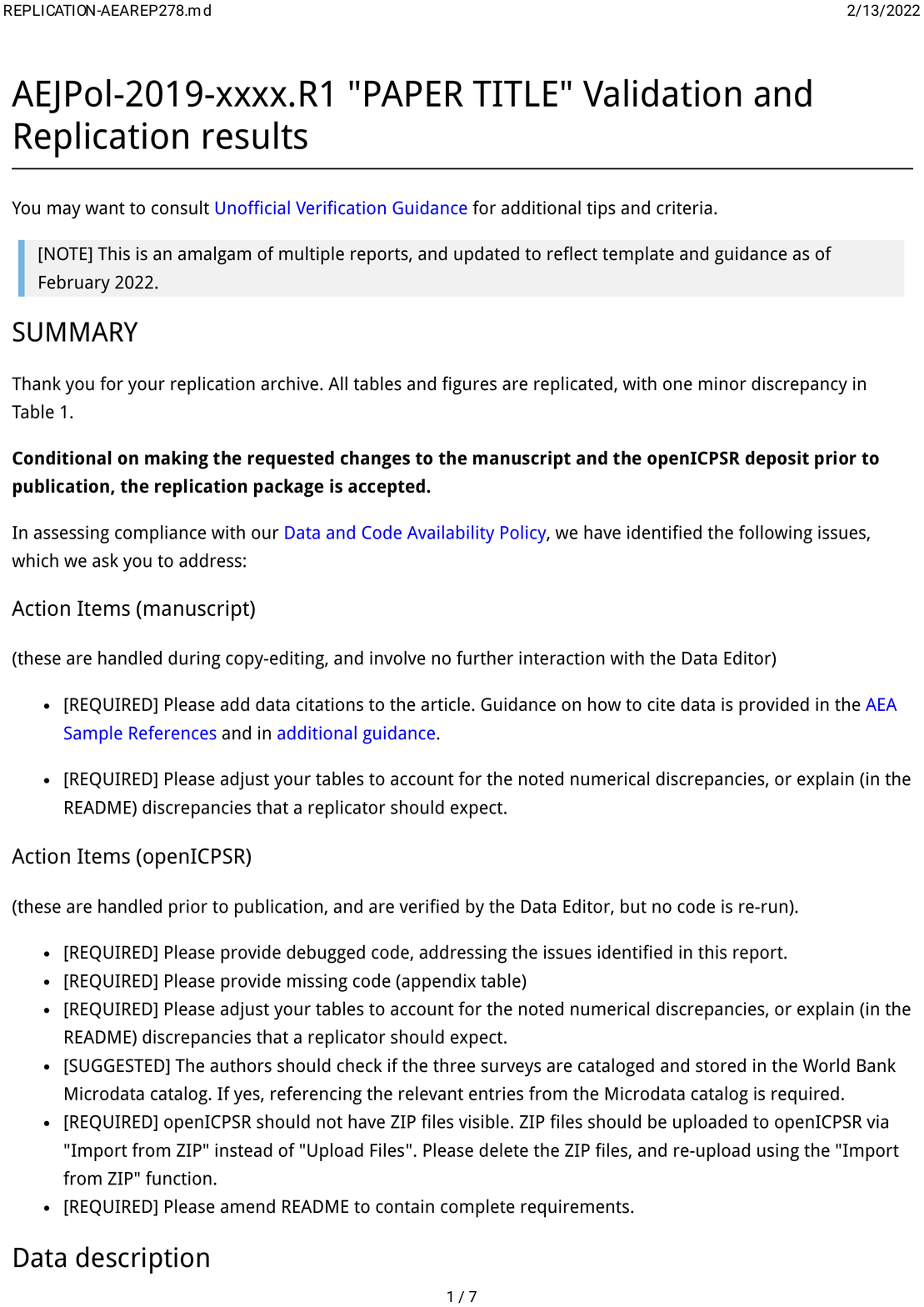}

\includegraphics[page=2,width=0.9\textwidth]{REPLICATION-AEAREP278.pdf}

\includegraphics[page=3,width=0.9\textwidth]{REPLICATION-AEAREP278.pdf}

\includegraphics[page=4,width=0.9\textwidth]{REPLICATION-AEAREP278.pdf}

\includegraphics[page=5,width=0.9\textwidth]{REPLICATION-AEAREP278.pdf}

\includegraphics[page=6,width=0.9\textwidth]{REPLICATION-AEAREP278.pdf}

\includegraphics[page=7,width=0.9\textwidth]{REPLICATION-AEAREP278.pdf}

\end{document}